\documentclass[12pt]{iopart} \usepackage{graphicx}

\expandafter\let\csname equation*\endcsname\relax
\expandafter\let\csname endequation*\endcsname\relax
\usepackage{amsmath,amssymb}
\usepackage{caption}

\usepackage{epstopdf} \usepackage{tabularx} \usepackage{caption}
\usepackage{bm} \usepackage{array}

\usepackage{color,footmisc} 

\begin{document}

\title[Vacuum flux surfaces by tilted coils]
{Large vacuum flux surfaces generated by tilted planar coils}

\author{Jessica L.~Li\footnote{Present address: Princeton University,
    Princeton, NJ 08544, USA\label{footPrinc}}, Jacob Austin,
  Kenneth C.~Hammond\footnote{Present address: Max Planck Institut f\"ur
     Plasmaphysik, D-17491 Greifswald, Germany},
  Ben Y.~Israeli\footref{footPrinc},
  Francesco A.~Volpe}

\address{Department of Applied Physics and Applied Mathematics,
  Columbia University\\ Mail Code: 4701, New York, NY
  10027, USA} 
\eads{\mailto{fvolpe@columbia.edu}}
\vspace{10pt}
\begin{indented}
\item[]
\end{indented}

\begin{abstract}
Helical equilibria can be generated by 
arrangements of planar coils similar to tokamaks, but without a 
central solenoid and with the toroidal field (TF) coils tilted with 
respect to the vertical.
This is known from earlier numerical works, e.g.~P.E.~Moroz, {\em Phys.~Plasmas} {\bf 2}, 4269 (1995). 
        However, such concept tends to need large 
coils (of low aspect ratio) but form small plasmas (of large aspect ratio). 
Here it is numerically shown that larger, more attractive vacuum flux surfaces 
-relative to the size of the device- can be generated by carefully optimizing 
the inclination of the TF coils and currents in the various coil-sets.
Vacuum configurations 
of aspect ratios as low as 4 are found for 6 tilted TF circular coils.
Higher numbers of TF coils have advantages (smaller effective ripple)
and disadvantages (lower rotational transform, smaller plasma). Finally, the
aspect-ratio $A$ of the vacuum flux surfaces is quantified as a function
of the ratio $A_c$ of the coil-radius to the radial location of the
coil-center. It is found that, 
in order to minimize $A$, it is beneficial to interlink or
marginally interlink the TF coils ($A_c \lesssim 1$).
\end{abstract}


\section{Introduction}
%
Modular coils in modern stellarators are characterized by complex
shapes. In parallel with the development of faster and cheaper
construction techniques, it is desirable that coil-shapes be
simplified while fulfilling all other stellarator optimization criteria. These
include but are not limited to
minimized neoclassical and turbulent
transport, improved energetic particle confinement, good
ballooning stability, etc. Recent works in coil-simplification 
include the design of modular coils that are planar on their outboard side,
to ease maintenance access and blanket-module replacement \cite{Gates}.
In another study, the REGCOIL code enabled coil-designs of reduced
curvature (thus simpler to build, and subject to reduced electromagnetic
stresses) compared to other numerical techniques,
but generating the same magnetic field \cite{REGCOIL}. 

Heliotrons/torsatrons, on the other hand, feature helical coils as
wide as $2R+2b$, rather than $2b$. Here $R$ is the major radius of the
plasma and $b$ is a quantity comparable with the minor radius $a$ of
the plasma, but larger, to allow space for the blanket. Note that in a
reactor $R \simeq $ 8-20 m (depending on the design) and $b \simeq$ 3m
\cite{Sagara10}.

In brief, it would be desirable for optimized stellarator coils to be simpler,
for example more planar, and for heliotron/torsatron coils to also be simpler, 
and more compact.

The heliac meets such criteria: it features circular
toroidal field (TF) coils of diameter $2b$ and only one large circular
coil of diameter $2R$, in the midplane of the device.
The TF coils are vertically
oriented, but non-axisymmetrically arranged according to a helical
magnetic axis. However, heliac experiments such as H-1
\cite{Blackwell03} and TJ-II \cite{Sanchez11} exhibit reduced
confinement compared to other helical devices of comparable size. This
is exemplified by the lower multiplying factor, $f_{ren}=0.25\pm0.04$,
for TJ-II in the International Stellarator Scaling \cite{yamada05}. 

While in the heliac the helical axis is generated by helically
{\em displacing} the coils, in another class of helical devices 
the TF circular coils are {\em tilted}. 

In the present paper, after briefly reviewing such devices
(Sec.~\ref{subsec:rev}),
we lay out the motivation for investigating a particular sub-set in which
the coils are planar, tilted with respect to the vertical, 
interlinked to each other, and the plasma current is negligible
(Sec.~\ref{subsec:Motiv}). In particular, we consider  
configurations with $N$=3-18 tilted TF coils. We optimize
them for maximum plasma volume or, equivalently, minimum aspect ratio,
as a function of the TF coils' inclination and of the currents in the
TF and poloidal field coil-sets. The rationale for such optimization is that
the aspect ratio $A$ must be reasonably low for a stellarator reactor
to be attractive. 
Indeed, after explaining the principle of rotational transform by tilted coils
in Sec.~\ref{sec:Principle} and describing the numerical method in
Sec.~\ref{sec:NumMeth}, we obtain values as low as $A$=4
in Sec.~\ref{sec:NumRes}.
In Sec.~\ref{sec:NumRes} we also analyze 
the dependence of $A$ and of the profiles of
effective helical ripple $\epsilon_{eff}$ and rotational transform 
$\iota$ upon the number of coils $N$, coil-tilt $\theta$
and normalized coil location $A_c$, defined in Sec.~\ref{sec:NumMeth}. 
 
This is the first extensive optimization of this nature. To enable
high-resolution scans of the large, multi-dimensional parameter space,
only {\em vacuum} flux surfaces were computed in the present study.
These can be considered low-beta approximations of plasma equilibria.

Equilibrium calculations at finite beta go beyond the scope
of the present paper and are left as future work, but are expected to reveal
even larger plasmas (lower aspect ratios), thanks to their finite bootstrap
current. More generally, finite plasma currents are known to lead to
larger plasma volumes \cite{Clark14} and, of course, higher rotational
transform. Similar ideas underpinned the
NCSX modular-coil quasi-axisymmetric stellarator design: the concept, 
since renamed QUASAR, self-consistently took 
advantage of finite boostrap-current to assist in generating
rotational transform and confine large plasmas of low aspect ratio
\cite{Zarn01,Ware06}.

\section{Background and motivation} 
\subsection{Brief review of tilted coil devices} \label{subsec:rev}

In the first work of this kind \cite{Georgi74}
planar coils were tilted both around the vertical and around the ``non-trivial''
horizontal axis (the trivial axis being the axis of symmetry of the
circular coil). The locus of the coil-centers was a circle. In another
concept \cite{Reiman83} the coils were helically displaced
as in a heliac, but tilted and
non-circular. Other arrangements of planar coils that generate helical
fields can be found in Refs.~\cite{Popov66,Rehker73}. 

Starting in the late 1980's, variants
\cite{Bykov88,Todd90,Bykov91,Georgi93,Moroz_PoP95} of
Ref.~\cite{Georgi74} started receiving a great deal of
attention. These variants featured fewer TF coils than the original
idea (ranging between 2 \cite{Todd90} and 9 coils \cite{Moroz_PoP95},
instead of 24 \cite{Georgi74}). This allowed for simpler construction,
larger confined volume, higher rotational transform, but also a higher
degree of non-axisymmetry and more pronounced toroidal
ripples. Another difference is that the coils were only tilted with
respect to the vertical plane, i.e., around their non-trivial
horizontal axes.

As first noted in \cite{Bykov88,Bykov91},  
these configurations
are, in effect, heliotrons/torsatrons (currents flow in the same direction
in all coils, unlike classical stellarators, where they are alternated). 
Their helical coils 
have poloidal mode number $m=$1 like other heliotrons/torsatrons, but
toroidal mode number $n=$1 as well.
Due to this $m=n=1$ peculiarity, the ``helical'' coils are, in fact, circular.
In contrast, most other heliotrons/torsatrons have $n=$5-10.

In some cases \cite{Bykov88,Todd90,Bykov91,Georgi93}
each TF coil was ``interlinked'' or ``interlocked'' to
every other TF coil, as if all TF coils had been pushed toward a central column
(similar to a spherical tokamak) 
and beyond. This resulted in coils of diameter $\sim2R$ comparable
to the device diamater, as is typical of heliotrons/torsatrons.
In other cases \cite{Bykov88,Bykov91,Georgi93,Moroz_PoP95} the TF coils
were not interlinked and were smaller, of diameter $\sim2b$, similar to a
heliac.

Starting in the mid-1990's, Moroz numerically investigated several
{\em non-interlinked} configurations 
\cite{Moroz_PoP95,Moroz_FST96,Moroz_PoP96,Moroz_PLA97,Moroz_PPCF97,Moroz_NF97,Moroz_PPCF98,Moroz_PLA98}. In
some of them the coils were planar
\cite{Moroz_PoP95,Moroz_FST96,Moroz_PPCF97}, including non-circular
shapes \cite{Moroz_PoP95,Moroz_PPCF97}. In others they were non-planar
\cite{Moroz_PoP96,Moroz_PLA97,Moroz_NF97,Moroz_PPCF98,Moroz_PLA98,Moroz_PRL96},
e.g.~helical on the inboard \cite{Moroz_PPCF98,Moroz_PLA98} or
outboard side \cite{Moroz_PLA97,Moroz_NF97}, as also proposed
\cite{Furth68,Boozer82,Cooper00} and experimentally realized
\cite{Oishi14} elsewhere.

Some studies assumed a net plasma current
$I_p$ \cite{Moroz_PoP96,Moroz_PLA97,Moroz_NF97,Moroz_PLA98}, others
assumed $I_p\simeq$0 \cite{Moroz_PoP95,Moroz_PRL96}, others still
compared cases with and without plasma current
\cite{Moroz_FST96,Moroz_PPCF98}.

Finally note that, along with 50 modular non-planar coils, W7-X 
is equipped with 20 tilted planar coils. Of these, 10 (of the ``PCA'' type) are
tilted in one direction with respect to the vertical, and the other 10
(of the ``PCB'' type) are tilted
in the opposite direction, and by a different amount \cite{Andreeva15}.

\subsection{Tilted, interlinked, planar coil torsatrons}
\label{subsec:Motiv}

In the present paper we argue that {\em planar, interlinked} coils
configurations with $I_p=0$
are particularly appealing, and we numerically optimize them for maximum
plasma size (in a low $\beta$, vacuum limit).

{\em Planar} coils are obviously appealing from a
manufacturing point of view, because they are simpler to construct than
other stellarator coils. 

In most of the paper we restrict to {\em interlinked} coils, partly because of
their relevance to the CNT and CIRCUS experiments at Columbia University
and partly because, as it will be shown in Sec.~\ref{subsec:depAc},
interlinked and marginally interlinked coils yield larger plasmas,
of lower aspect ratio. 

CNT is equipped with just two interlinked coils and two poloidal field or
vertical field (VF) coils. CNT was the first device to toroidally
confine non-neutral plasmas \cite{pedersen2004,kremer2006,brenner2008}
and plasmas with various degrees of quasi-netrality
\cite{sarasola12}. Its focus has recently shifted to 3D diagnostic  
image inversion \cite{hammond_rsi2016}, error 
fields \cite{hammond2016}, high beta
\cite{hammond_pop2017} and overdense microwave heating \cite{hammond_overd} 
in neutral stellarator plasmas.

CIRCUS \cite{Clark14} is 
equipped with six interlinked TF coils of adjustable tilt
$\theta=40-60^\circ$ with respect to vertical, and two up-down
symmetric pairs of poloidal field coils, denominated respectively VF
and quadrupole field (QF) coils. CIRCUS aims at experimentally
generating or amplifying rotational transform $\iota$ by using more than two 
tilted planar coils, for the first time.
Here ``generating'' refers to generating a finite
$\iota$ and creating flux-surfaces, even in the absence of plasma
current ($I_p$=0). ``Amplifying'' $\iota$ by means of tilted coils  
refers to obtaining a higher $\iota$ than if the coils were not tilted;
a finite $\iota$, however, is necessary to begin with (this could be an
external rotational transform from non-axisymmetric coils,
or could be due to $I_p\ne$0). 
CIRCUS was originally conceived as a tokamak-torsatron hybrid in which a finite 
$I_p$ generates a finite $\iota$ and the tilted coils increase
or {\em amplify} it (as if they imparted ``kicks'' to the helical field-lines,
and thus twisted them even more)  \cite{Clark14}. In the present paper,
however, it is predicted that CIRCUS can operate as a pure torsatron as well
and {\em generate} $\iota$ even in the absence of finite $I_p$.

This is an intermediate step toward even higher numbers $N$ of tilted TF coils.
High $N$ are attractive for tokamak-torsatron hybrids ($I_p\ne$0):
compared with equivalent tokamaks adopting the same $N$ (say, $N$=18), these
hybrids are 
expected to generate more rotational transform in spite of
requiring a 25-50\% lower plasma current $I_p$ \cite{Clark14,Spong13},
making disruptions less likely and less harmful. 
Incidentally, it is well-known 
that small fractions of external rotational transform
dramatically reduce the disruptivity of tokamak and
hybrid plasmas \cite{Grieger85,Pandya15}. 
Hence, plasmas confined by $N$=18 tilted coils
are expected to be significantly less disruptive than equivalent tokamak
plasmas.
In addition, hybrid tokamak-torsatron plasmas of high $N$ are
more axisymmetric than equivalent tokamaks or torsatrons:
the effective helical ripple $\epsilon_{eff}$ is expected to be even smaller
than in equivalent tokamak plasmas \cite{Clark14,Spong13} and much smaller
than in typical stellarator and heliotron/torsatron plasmas, with
benefits for confinement. 

Finally, we restrict to {\em current-free}
configurations that do not require a solenoid nor current drive, due
to their attractiveness for steady state at high plasma
density. Note that most current drive mechanisms tend to be
inefficient at the high densities encountered in the high-density
H-mode at W7-AS \cite{McCormick02}.

\section{Physical principle of rotational transform generation by
  tilted planar coils} \label{sec:Principle}

\subsection{Interlinked coils}

\begin{figure}[t]
  \begin{center}
    \includegraphics{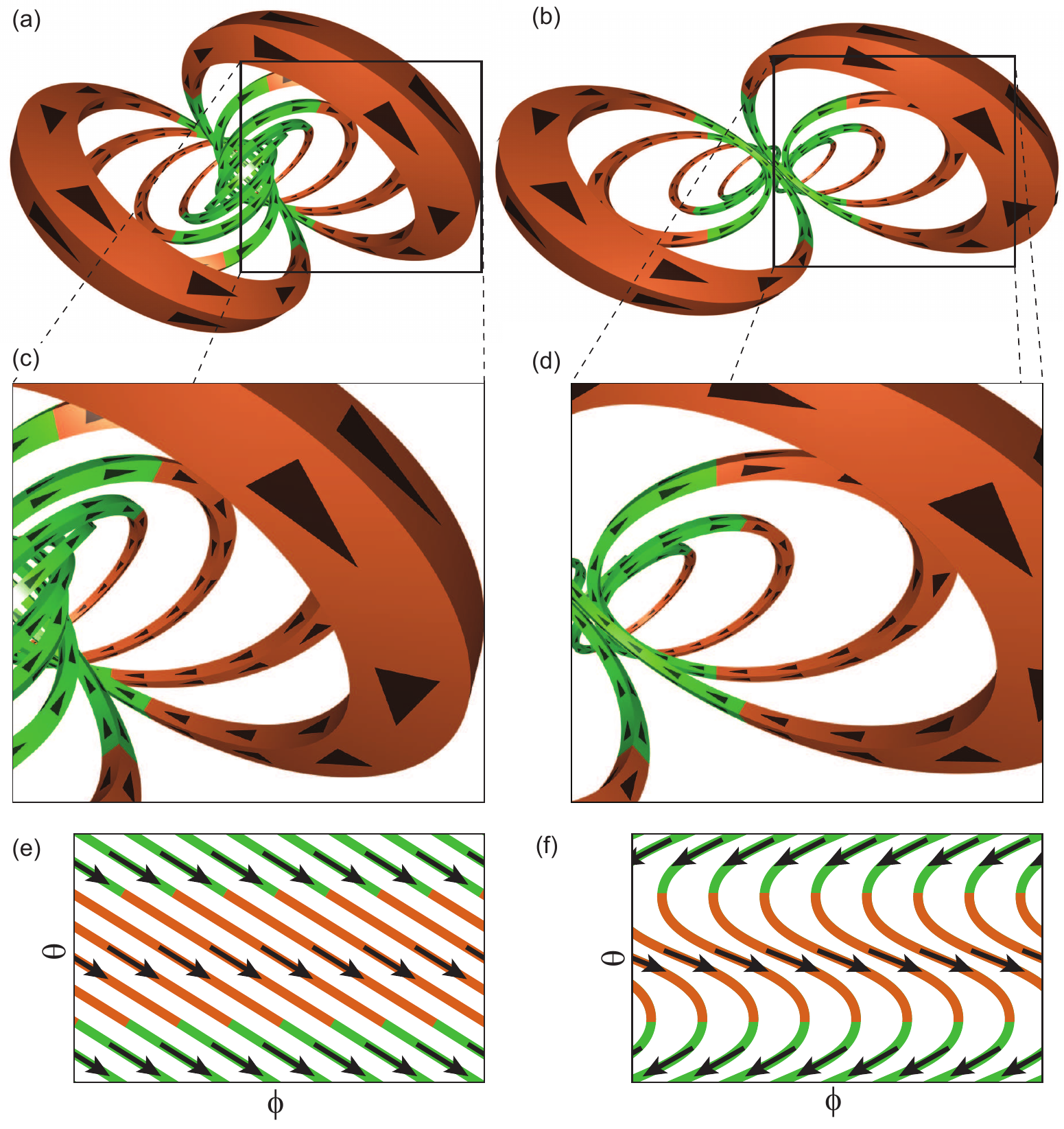}   
    \caption{Computer-rendered view of a set of
      (a) interlinked and (b) not interlinked toroidal field (TF) coils,
      all tilted by the same angle $\theta=45^\circ$ relative to vertical.
      The inboard and outboard side of each coil (located respectively
      at smaller and larger major radii $R$) are colored in green and orange.
      Arrows denote the verse of the currents in the coils. (c)-(d): details
      of (a)-(b). (e)-(f): corresponding ``unwrapped'' coil-winding surfaces,
      with coil-current patterns plotted as functions of poloidal and toroidal
      angle. 
    }
    \label{fig:Cartoon}
  \end{center}
\end{figure}

When the tilted coils are interlocked, it is intuitive that they are
equivalent to a torsatron. To visualize this, imagine being a local observer at 
toroidal location $\phi$ inside the torus.
Let us call ``outer'' and ``inner'' the regions at larger and smaller
major radii $R$, respectively,
and color-code them in orange and green in Fig.~\ref{fig:Cartoon}a. 
The observer in $\phi$ will only see the ``outer part'' (orange) of coils
centered at nearby locations (say, in a range $\phi\pm\pi/2$), 
and only the ``inner part'' (green) of 
nearly diametrically opposite coils, located at $\phi+\pi\pm\pi/2$.
The consequence is illustrated in Fig.~\ref{fig:Cartoon}c:
the local observer has the perception of {\em helical} windings,
all carrying current in the same direction 
(poloidally clockwise and toroidally counter-clockwise -that is, ``pointing away from the observer''- in the specific example pictured).
This is because if one takes poloidal
cross-sections at incremental toroidal angles (not shown),
the cross-sections of the coils will rotate in a definite poloidal direction. 

Equivalently, the ``unwrapped'' coil-winding surface looks like in 
Fig.~\ref{fig:Cartoon}e: all coil-currents have the same helicity everywhere.

\subsection{Non-interlinked coils}

In the case of non-interlinked coils (Fig.~\ref{fig:Cartoon}b),  
the observer in $\phi$ only sees local coils
(at toroidal locations $\phi \pm \pi/2$ or closer), but not the remote ones.
Now consider the poloidal cross section of a single tilted coil. 
This intersects a vertical plane at two locations.
At those two locations, the current obviously flows in opposite directions. 
If one now considers several coils, all tilted in the same direction, and
takes poloidal cross-sections at incremental
toroidal angles, all coil cross-sections will move upward or all downward.
This is not a helical device, where all coil cross-sections move
poloidally clockwise, or counter-clockwise. 
More specifically, this is neither a torsatron (where all currents
point in the same helical direction and verse) nor a classical
stellarator (where adjacent coil-currents have alternate verses).
Rather, some currents ``point toward the observer''
(see top left of Fig.~\ref{fig:Cartoon}d). Their helicity is inconsistent
with the other currents in Fig.~\ref{fig:Cartoon}d.  
Equivalently, the unwrapped coil-winding surface looks like in 
Fig.~\ref{fig:Cartoon}f: each TF coil contributes currents of a certain
helicity on the outboard side (orange) and of opposite helicity on the inboard
side (green). 

The
key, however, is that (1) these coils generate a helical magnetic axis and (2)
the plasma column rotates and changes shape with $\phi$. Both features (1) and
(2) were noticeable in Fig.~2 of Ref.~\cite{Moroz_PoP95}. Incidentally,
that figure referred to $\phi=$0-0.35 in a device with interlinked coils,
but is easily generalized to $\phi=$0-0.7  
by stellarator symmetry. Features (1) and (2) will also be
visible in Fig.~\ref{fig:FluxSurf} of the present article, also for interlinked
coils. 
Points (1) and (2) are two of
the three sufficient conditions to generate helical
transform, the third one being finite plasma current 
\cite{Spitzer58,Helander14}. Biot-Savart calculations confirm the generation of 
helical fields, even when the coils are not interlinked
\cite{Bykov88,Bykov91,Georgi93,Moroz_PoP95}
and codes confirm the existence of equilibria \cite{Moroz_FST96,Moroz_PoP96,Moroz_PLA97,Moroz_PPCF97,Moroz_NF97,Moroz_PPCF98,Moroz_PLA98}.

\subsection{Coil-tilt always amplifies rotational transform, but only
generates it under special circumstances}

Note that arbitrary sets of poloidal field coils and tilted TF coils
(whether interlinked or not), energized with arbitrary currents, only
generate infinitesimal vacuum flux surfaces, or none at all. They
would still ``amplify'' $\iota$, in the sense that, if
field-lines are already twisted by other means (Ohmic plasma current,
current drive, effect of bootstrap current, external rotational
transform), tilted planar coils can give them further ``kicks'' and
twist them even more. However, in order for these
configurations to act as ``sources'' or ``generators'' of rotational
transform, the TF coil inclination and the coil-currents must be
properly chosen, as it will be shown in Sec.~\ref{sec:NumRes}. 

Finally, because
all TF coils are tilted in the same direction and energized in the
same direction, they generate a net vertical field similar to
heliotrons/torsatrons and unlike stellarators, calling for 
compensation by VF coils.

\subsection{Alternative point of view}     \label{subsec:Altern}
  
Consider the volume enclosed by tilted TF coils, whether
interlinked (Fig.~\ref{fig:Cartoon}a) or not (Fig.~\ref{fig:Cartoon}b).
To clarify, in the case of the coils being interlinked,
the ``inboard side'' of the volume of interest (lying at smaller major radii)
is bound by the coils' ``outside''  
(the side facing larger minor radii of the coils). 

Consider now an arbitrary location within this volume.
In that location, all the tilted TF coils, whether close
or diametrically opposite, generate toroidal fields of
the same sign. The same is true in any 
other arbitrary location. That is, the sign of the toroidal field
$B_\phi$ is uniform. 
The vertical field $B_z$ from the TF coils, on the other hand,
is sheared in the major radius direction,
and changes sign near the inboard wall.

The VF and, to some extent, the QF coils superimpose 
an additional vertical field, nearly uniform. This moves the magnetic null
(roughly the magnetic axis) to outer radii.

The QF and, to some extent, the VF coils add a radial field $B_R$ 
that is vertically sheared, and changes sign at the midplane of the device.

These vertically sheared $B_R(z)$ and radially sheared $B_z(R)$  
combine to create a poloidal field $B_\theta(r)$, where $r$ is the minor radius.
The latter, in
combination with $B_\phi$, creates nested flux surfaces with
rotational transform.

As for the toroidal dependence, both $B_R$ and $B_z$ oscillate with
period $2\pi/N$ in direction $\phi$, but out of phase with each other.
This results in a helical magnetic axis, also generating rotational
transform. Note that, for higher $N$, the oscillations become more frequent 
but also smaller in amplitude, to the detriment of rotational transform
(which might partly explain why the plasma becomes smaller). 
For $N\rightarrow\infty$, the magnetic axis is perfectly axisymmetric. 

The considerations made in the present Section,
\ref{subsec:Altern}, apply equally to interlinked and non-interlinked
configurations.

\section{Numerical method}    \label{sec:NumMeth}

One of the goals of this study is to minimize the plasma aspect-ratio as
a function of the TF coil-tilt and of the TF, VF and QF coil-currents or,
equivalently, of coil-current ratios. Two such ratios suffice, because the 
goal is to
maximize the plasma volume or, equivalently, minimize the aspect ratio.
Therefore, the field topology is important, but the field magnitude is not,
and is defined on the net of a scaling factor. 

For this reason, for each combination of current-ratios and tilts we 
identified the Last Closed Flux Surface (LCFS) by means of a field-line tracer
and computed the volume $V$ of the
enclosed plasma and the toroidally averaged
major radius $R$ of the magnetic axis. 
From these pieces of information we
deduced the minor radius of the plasma,
$a=\sqrt{V/2\pi^2R}$ and, ultimately, its aspect ratio $A=R/a$.  

The field-line tracer used was FIELDLINES \cite{Lazerson_NF16}.
As usual this was interfaced to the MAKEGRID Biot-Savart code, but 
with {\em ad hoc} modifications. 
Namely, normally the code discretizes the current-carrying coils in
short current filaments and numerically integrates the Biot-Savart law
to compute the magnetic field in a location of interest. 
This is appropriate for complicated 3D coils. Here, however, 
similarly to Moroz in his UBFIELD field-line tracing code \cite{Moroz_PoP95},  
we took advantage of the coils being circular and the generated
field being known analytically \cite{Garrett63,Jackson,Simpson01}:
\begin{eqnarray}
  B_r & = &
  \frac{\mu_0 I}{2\pi}
  \frac{z}{\alpha^2\beta r}
  \lbrack(a^2 + r^2 + z^2) E(k^2) - \alpha^2 K(k^2)\rbrack,
  \\
  B_z & = &
  \frac{\mu_0 I}{2\pi}
  \frac{1}{\alpha^2\beta}
  \lbrack(a^2 - r^2 - z^2) E(k^2) + \alpha^2 K(k^2)\rbrack.
\end{eqnarray}

Here $r$ and $z$ are cylindrical coordinates relative to the coil center, 
$K$ and $E$ are complete elliptic integrals of the
first and second kind, respectively \cite{Abramowitz}
and 
\begin{eqnarray}
	\alpha^2 & = & a^2+r^2+z^2-2ar,
\\
	\beta^2 & = & a^2+r^2+z^2+2ar,
\\
	k^2 & = & 1-\frac{\alpha^2}{\beta^2} = \frac{4ar}{\beta^2}.
\end{eqnarray}

\begin{figure}[t]
  \begin{center}
    \includegraphics[width=8cm]{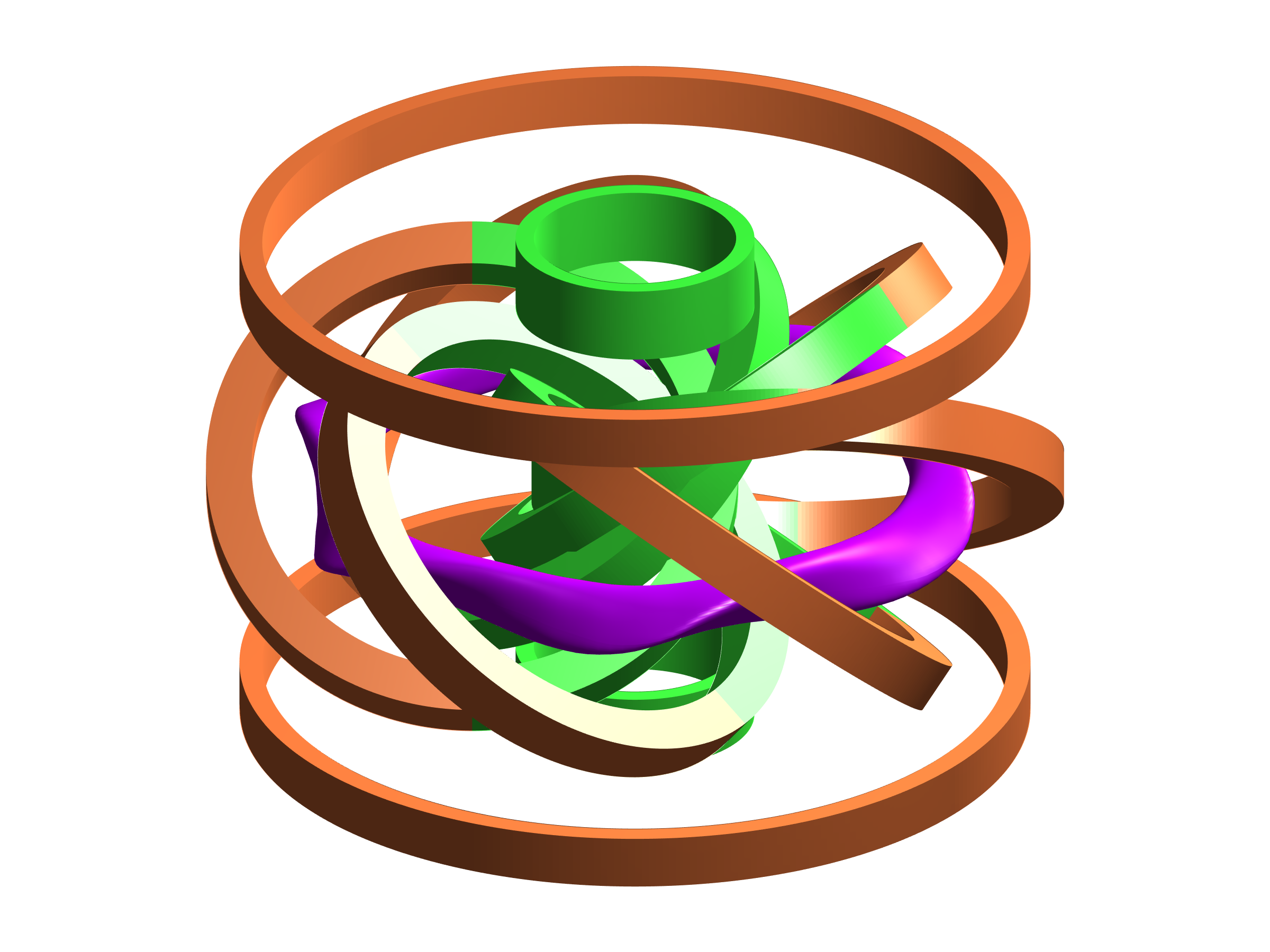}
    \caption{Rendering of plasma, vertical field, quadrupole
      field and $N$=6 tilted toroidal field
      coils in the CIRCUS device \cite{Clark14}, with high- and low-field side
      in green and orange as in Fig.~\ref{fig:Cartoon}. 
    }
    \label{fig:3DRend}
  \end{center}
\end{figure}

All coils in the present paper were modeled as finite-width arrays of the
circular filaments just described. The cross-sections (length $\times$
radial width)
are 3.5$\times$3.0 cm for the TF coils, 3.4$\times$1.5 cm for the VF coils,
and 5.0$\times$1.2 cm for the QF coils. The size and relative position
of the coils and the plasma is illustrated in Fig.~\ref{fig:3DRend}
in the case of CIRCUS, featuring $N=$6 tilted coils.
Further details can be found in Ref.~\cite{Clark14}.

This semi-analyitic approach sped up the calculations for a single coil
configuration by nearly two orders of magnitude, which allowed
investigating more configurations in the same amount of time. This resulted in
broad, fine scans of the parameter space. 
In particular, for each choice of the number of TF coils, $N$, 
and their tilt angle with respect to the vertical, $\theta$, 
we numerically scanned the coil-current ratios  
$I_{\text{TF}}$/$I_{\text{VF}}$ and $I_{\text{QF}}$/$I_{\text{VF}}$.

Various field-lines were traced for each combination of $N$, $\theta$
and $A_c$. In the present article all TF coils are tilted by the same
angle $\theta$; we do not consider the case in which they could be tilted
by different amounts. 
The quantity $A_c$ is defined as the ratio between the major radius $R_c$
at which the TF coils are centered, and the radius $a_c$ of the TF coils. 
This 'normalized coil location' is a measure of how interlinked the coils
are (interlinked for $A_c<1$, not interlinked for $A_c>1$). We will
sometimes refer to it as 'coil aspect ratio' (not to be confused with
the ratio of the coil radius to the coil half-thickness).
It is necessarily lower than the plasma aspect ratio $A$, although ideally it
should not be much lower: $A_c \ll A$ means that the TF coils are much larger
than the poloidal cross-section of the plasma.

Field-lines were traced with such 
a numerical tolerance that, even after hundreds of toroidal turns,
they are uncertain to less than $\pm$1mm. 

The LCFS was identified as the 
outermost laminar surface outside of which field-lines are open and reach
the boundary of the computational domain. Such identification took place
in two parts: a 30-step coarse scan to isolate a promising radial interval, 
followed by a 30-step fine scan in that interval.
This can yield a precision of up to one part in 900 at the cost of tracing
just 60 field-lines (``up to'' because
for redundancy the interval examined in the fine
scan was wider than the interval identified in the coarse scan).
The idea is easily generalized to a bisection method (repetitive 2-step scans,
zooming more and more on the LCFS).

\section{Numerical results}
\label{sec:NumRes}

\subsection{6-coil configuration}

\begin{figure}[t]
  \begin{center}
    \includegraphics{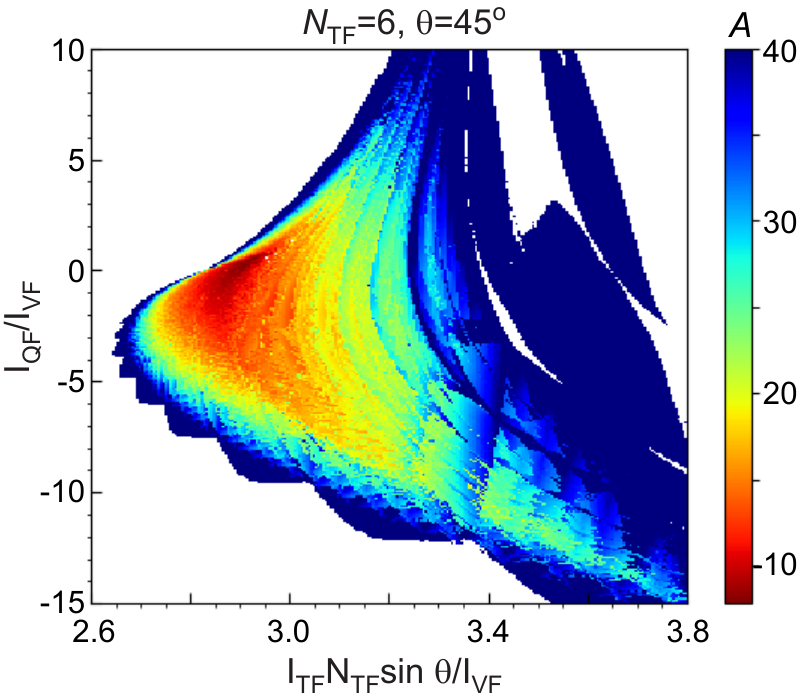}
    \caption{Contours of plasma aspect ratio $A$ as a function of coil
      current-ratios in a configuration of 2 QF and 2 VF
      coils, as well as $N$=6 TF coils of
      normalized radial location $A_c=$0.67, tilted by
      $\theta=45^\circ$ with respect to the vertical.}
    \label{fig:Cont}
  \end{center}
\end{figure}

The CIRCUS device \cite{Clark14}
features six TF coils of inclination adjustable in the range
$\theta = 40-60 ^\circ$, relative to the
vertical. Each TF,
VF and QF coil consists of 69, 54 and 56 turns, respectively.
$I_{TF}$, $I_{VF}$ and $I_{QF}$ denote the total currents in ampere-turn At,
not in A. Fig.~\ref{fig:Cont} presents the plasma aspect ratio $A$
as a function of the coil current-ratios, for $\theta=45^\circ$.
$I_{TF}$ is replaced
by $N I_{TF} \sin \theta$ to isolate the vertical field component generated
by the TF coils and multiply it by the number of coils. 

\begin{figure}[t]
  \begin{center}
    \includegraphics{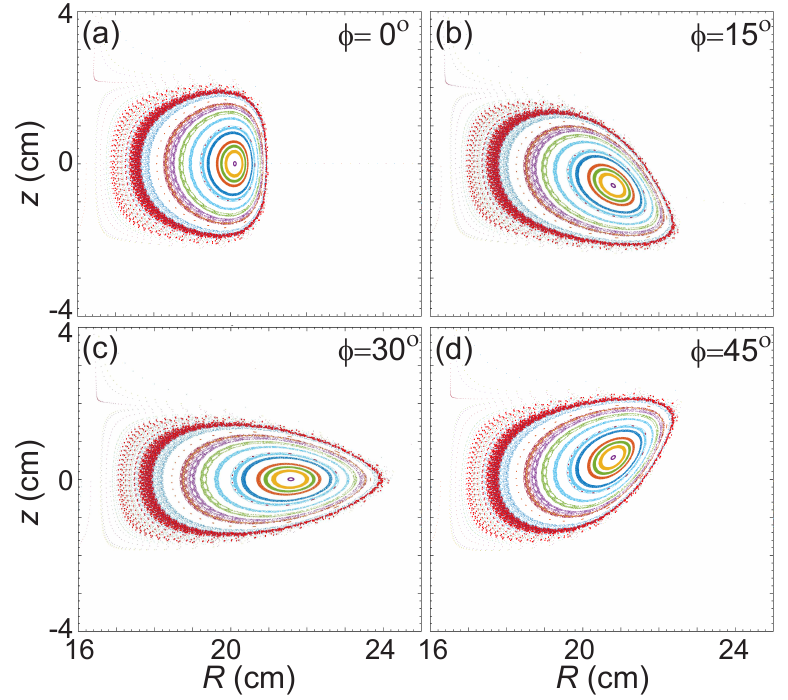}
    \caption{Poincar\'e plots of flux-surfaces in optimal configuration
      (of lowest $A$) from Fig.~\ref{fig:Cont}, for a choice of the major
      radius location of the magnetic axis corresponding to the
      CIRCUS device \cite{Clark14}. The poloidal cross-sections
      are taken at four toroidal locations
    corresponding to the beginning, $1/4$, $1/2$ and $3/4$ of a field-period.}
    \label{fig:FluxSurf}
  \end{center}
\end{figure}

In this as well as in Figs.~\ref{fig:FluxSurf}-\ref{fig:TiltDependProfiles},
the normalized radial location 
of the TF coils was set to $A_c=$0.67, as in CIRCUS \cite{Clark14}. 

The lowest aspect ratio for this choice of $\theta$, $A=$7.8, 
is obtained for $I_{QF}/I_{VF}$=0.35 and $I_{TF}/I_{VF}=-0.69$.
The corresponding flux surfaces are plotted in Fig.~\ref{fig:FluxSurf} 
for the specific dimensions of the CIRCUS table-top device. 

Striations in Fig.~\ref{fig:Cont} and in similar contours in
Figs.~\ref{fig:NDepend}, \ref{fig:TiltDepend} and \ref{fig:AcDepend}
are due to rational surfaces near the LCFS. 

\subsection{Dependence on number of TF coils}   \label{sec:NDep}

\begin{figure}[t]
  \begin{center}
    \includegraphics{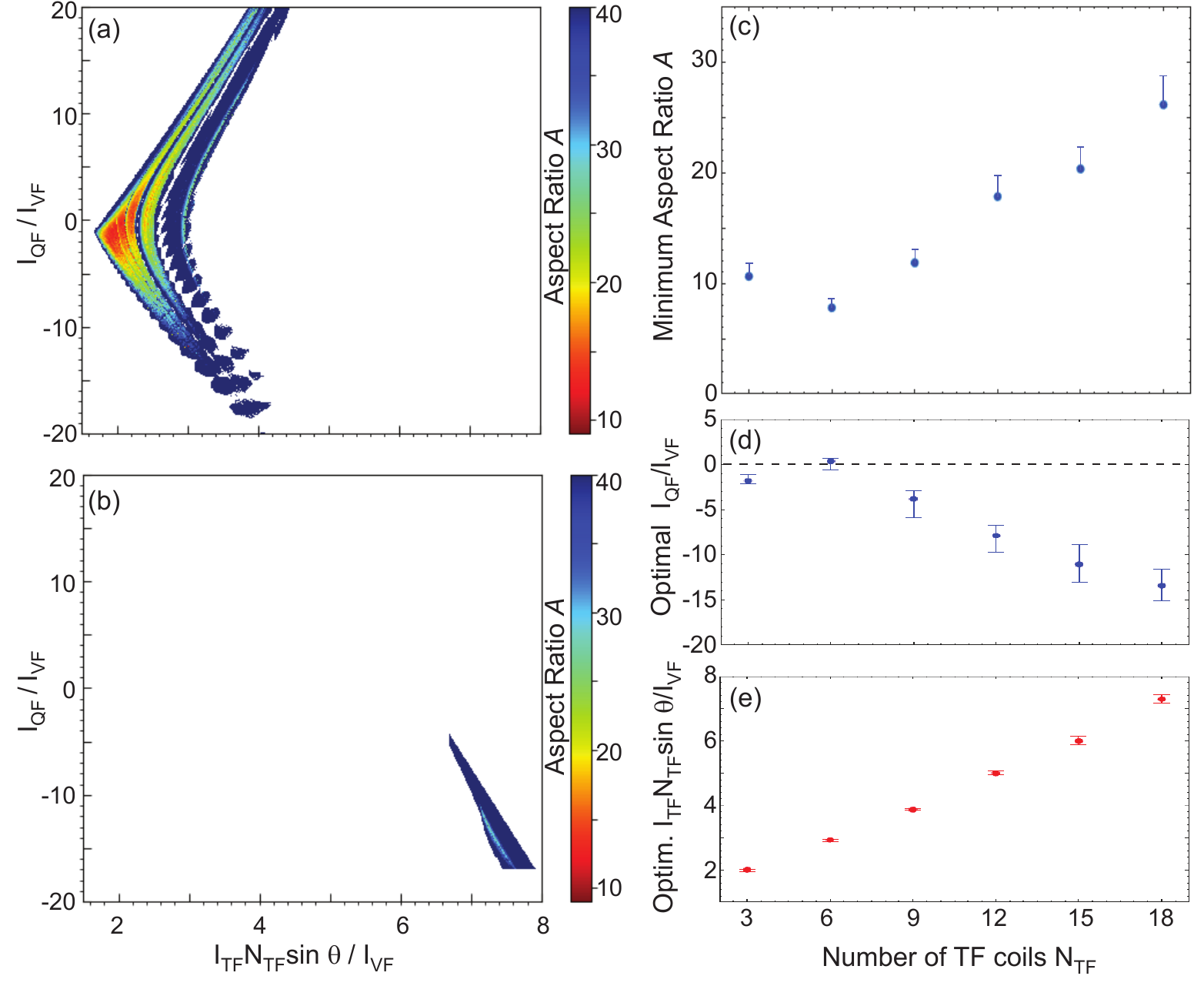} 
    \caption{(a)-(b) Like Fig.~\ref{fig:Cont} ($N$=6, $\theta=45^\circ$,
      $A_c$=0.67), but for 
      $N$=3 and $N$=18 tilted coils. 
      (c) Minimum plasma aspect ratio $A$ obtainable for various $N$  
      (for fixed $\theta=45^\circ$ and fixed VF and QF coil positions).
      (d-e) Current-ratios yielding optimal $A$, as functions of $N$.
      Also shown, in the form of error-bars, are: (c) ranges
      of near-optimal $A$ (within 10\% of the minimum) and (d-e) corresponding
      ranges of current-ratios yielding those values of $A$. This gives a
      measure of the sensitivity of $A$ to the coil-currents.}
    \label{fig:NDepend}
  \end{center}
\end{figure}

Numerical scans of the type presented in Fig.~\ref{fig:Cont} 
were performed for CIRCUS-like configurations with 
varying numbers $N$ of TF coils, all the rest remaining equal. 
Contours of $A$ are shown in Fig.~\ref{fig:NDepend}a-b for the
lowest and highest value considered, $N$=3 and $N$=18.
Contours for other values of $N$ are not shown for brevity, but the results of
the scan, in increments $\Delta N$=3, are summarized in Fig.~\ref{fig:NDepend}c:
the lowest $A$ is plotted for each $N$ as a function of the coil-current
ratios.
The tendency is for $A$ to increase with $N$. 
This is partly due to a decrease in ripple, leaving less space for the
plasma to ``bulge out'', which makes the plasma more axisymmetric,
but also smaller. The data point for $N=$3 is an outlier.
This could be due to the configuration being so non-axisymmetric, in that case,
that the VF generated by simple
circular (axisymmetric) VF and QF coils cannot effectively
balance the highly non-axisymmetric
VF generated by the few tilted TF coils. It is speculated that the
issue could be ameloriated by properly shaped non-circular VF and QF coils. 

It should be noted that the VF and QF coil-positions were kept constant.
It is possible that their optimization (for instance, by moving the coils
closer to the plasma as this gets
smaller and smaller) could have enlarged $V$ and reduced $A$, but at
increased computational cost. 

Flux-surfaces very similar to those shown in Fig.~\ref{fig:FluxSurf} were
obtained for different $N$, but are not shown for brevity. 
The main difference was that increasing $N$ resulted in smaller flux-surfaces
and, of course, shorter toroidal periods 360$^\circ/N$. 

\begin{figure}[t]
  \begin{center}
    \includegraphics{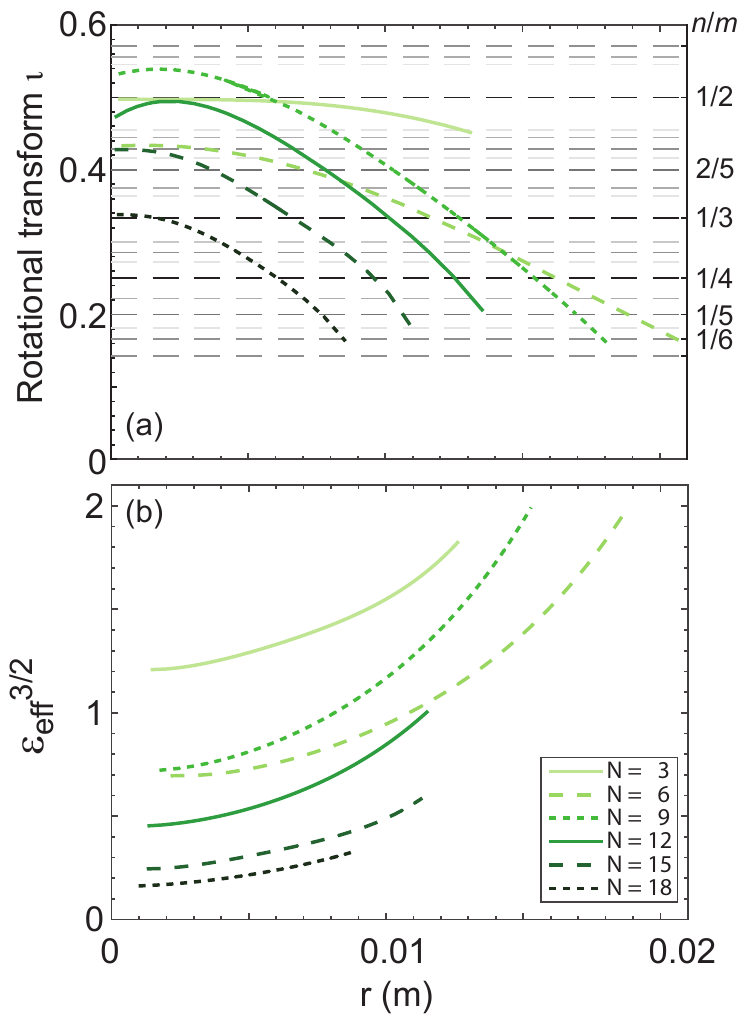}
    \caption{Radial profiles of (a) rotational transform and (b) effective
      ripple for various numbers of tilted TF coils, $N$, examined
      in Fig.~\ref{fig:NDepend}. To fix the ideas, we assumed coils
      of the same radii and radial locations as in the table-top CIRCUS device. 
      The horizontal dashed lines overlaid to the
      $\iota$ profile help localize rational surfaces and vacuum magnetic
      islands of $m \le$7. 
      }
    \label{fig:NDependProfiles}
  \end{center}
\end{figure}

Plotted in Fig.~\ref{fig:NDependProfiles} are 
the radial profiles of $\iota$ and $\epsilon_{eff}$, for various $N$.
The $\iota$ profiles were computed with FIELDLINES and found in agreement
with $\iota$ profiles from the equilibrium code VMEC \cite{VMEC}, 
used here at infinitesimal density and beta. 
The $\epsilon_{eff}$ profiles, instead, were computed using NEO \cite{NEO}.
The values of $\epsilon_{eff}$ are not as low as in a previous
paper dedicated to tokamak-torsatron hybrids ($I_p\ne 0$) with tilted coils
\cite{Clark14}. A possible explanation lies in the fact that
the torsatron plasmas discussed here ($I_p=0$) form at outer radii,
closer to the outboard coil boundaries, where ripples are more pronounced. 

Higher values of $N$ make the plasma more axisymmetric. The advantage is that
the effective helical ripple becomes smaller. The disadvantage is that the
vacuum rotational transform decreases as well, although it
remains acceptably high even at $N$=18 ($\iota=0.2-0.3$, comparable with
the earlier W7-AS). Note that the $\iota$ profile
peaks at the center, not at the edge. In this it differs from
typical stellarators and torsatrons, and is more similar to tokamaks. 
Also note that the magnetic shear is high and the $\iota$ profile crosses
several low-order rational values, $m/n$. Many magnetic islands can form
as a result, but all small (similar to the strategy of LHD and other heliotrons,
and opposite to the philosophy of the Wendelstein stellarator line).

\subsection{Dependence on coil tilt}

\begin{figure}[t]
  \begin{center}
    \includegraphics{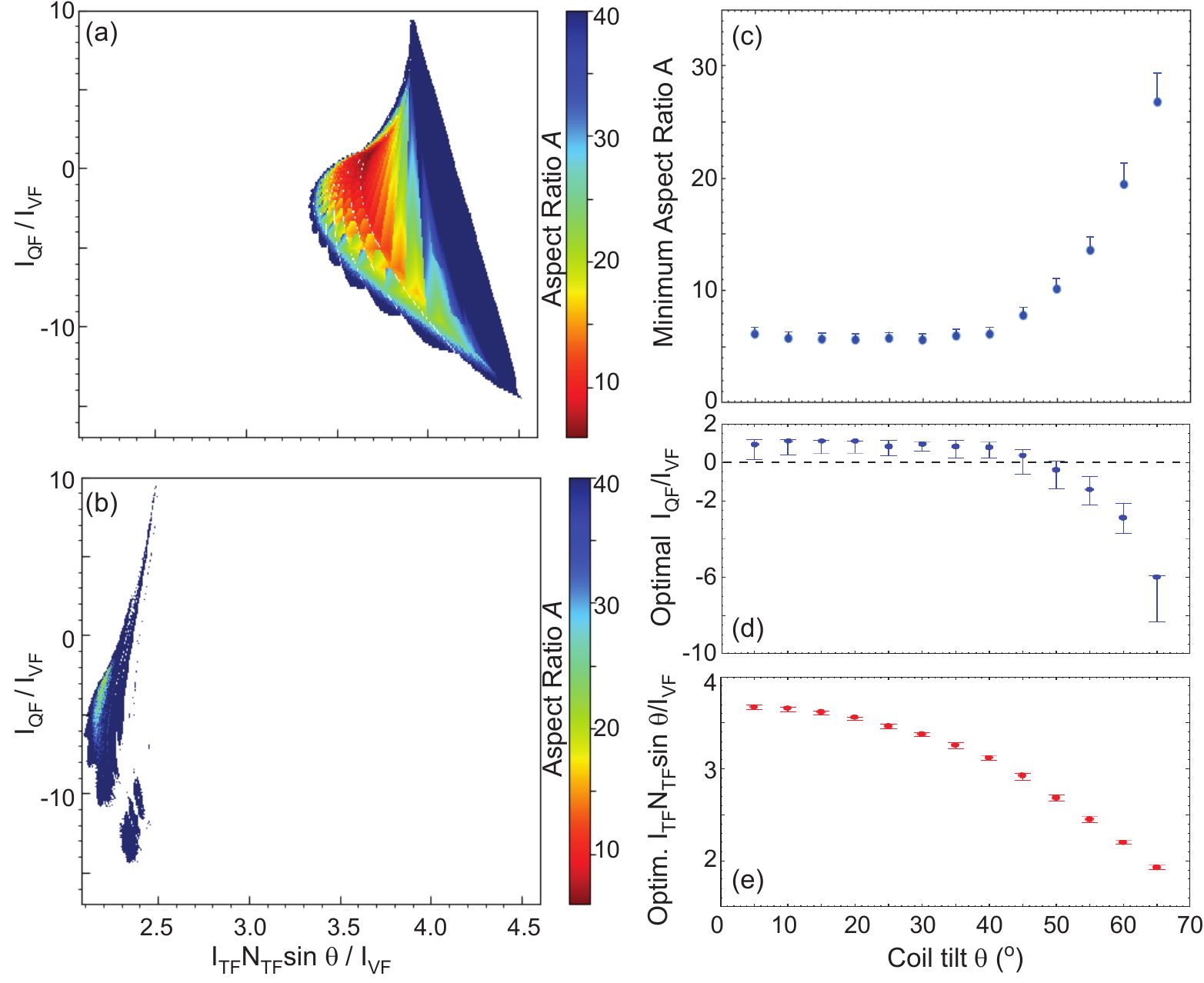} 
    \caption{(a)-(b) Like Fig.~\ref{fig:Cont} ($N$=6, $\theta=45^\circ$,
      $A_c$=0.67), but for 
      TF coil tilts $\theta=5^\circ$ and
      60$^\circ$ with respect to vertical.
      (c-e) Like Fig.~\ref{fig:NDepend}c-e, but as function of $\theta$. 
          }
    \label{fig:TiltDepend}
  \end{center}
\end{figure}

Next, numerical scans were performed for CIRCUS-like configurations
with 6 TF coils for tilt-angles varying from $\theta=5^\circ$
(Fig.~\ref{fig:TiltDepend}a) to
$\theta=60^\circ$ (Fig.~\ref{fig:TiltDepend}b) 
in steps of $5^\circ$, all the rest remaining equal.

The plot in Fig.~\ref{fig:TiltDepend}c exhibits a broad minimum of $A$
with respect to $\theta$, with the very minimum 
obtained at $\theta=30^\circ$. 

\begin{figure}[t]
  \begin{center}
    \includegraphics{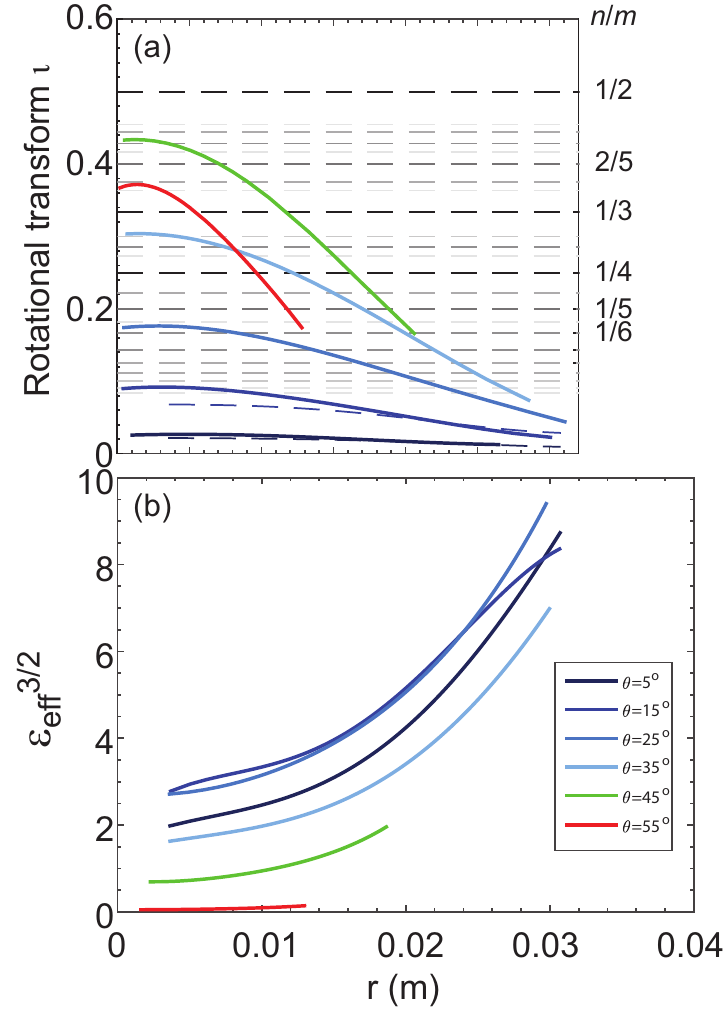} 
    \caption{Radial profiles of (a) rotational transform and (b) effective
      ripple for various TF coil-tilts $\theta$ examined
      in Fig.~\ref{fig:TiltDepend}. 
      }
    \label{fig:TiltDependProfiles}
  \end{center}
\end{figure}

For very small $\theta$, however, in spite of $A$ being attractively low,
the rotational transform is unattractively low
(Fig.~\ref{fig:TiltDependProfiles}a).  
This is because barely tilted coils are nearly indistinguishable from pure
TF coils: they only generate toroidal field and no rotational transform.

For large tilts ($\theta>45^\circ$), on the other hand,  
the field is nearly entirely vertical, and the torus
becomes oblated (basically, vertically ``squeezed''). As a result, the plasma
volume vanishes (Fig.~\ref{fig:TiltDepend}c). 

As noted in Sec.~\ref{sec:NDep},
optimizing the locations of the VF and QF coils instead of keeping them fixed
(e.g., moving them closer to smaller plasmas) 
could increase $V$ and reduce $A$. However, it would also 
increase the dimensionality of the scan and its computational cost. 

Plotted in Fig.~\ref{fig:TiltDependProfiles} are
the radial profiles of $\iota$ and $\epsilon_{eff}$, for various $\theta$.
For $\theta=5^\circ$ the TF coils are nearly vertical, similar to a tokamak.
Not surprisingly, the corresponding $\iota$ is very small. More tilted coils
impart higher rotational transform, reaching the maximum at about
$\theta=45^\circ$. Beyond that, $\iota$ decreases again, possibly due to the
plasma-shape oblation mentioned above. Higher tilts tend to yield
lower $\epsilon_{eff}^{3/2}$ as well. This is ascribed to the
field-line having less space to bulge out and deflect back in again.

\subsection{Dependence on coil aspect ratio}   \label{subsec:depAc}
The third parameter scanned is the normalized coil location or
{\em coil} aspect ratio, $A_c$.
As mentioned before, this is defined as the ratio between the major radius $R_c$
at which the TF coils are centered, and the radius $a_c$ of the TF coils.
Here we fix $a_c=0.16$m and scan $R_c$. 
Unlike $A$, this ratio can take values $A_c<1$, corresponding to the
coils being interlinked.

\begin{figure}[t]
  \begin{center}
    \includegraphics{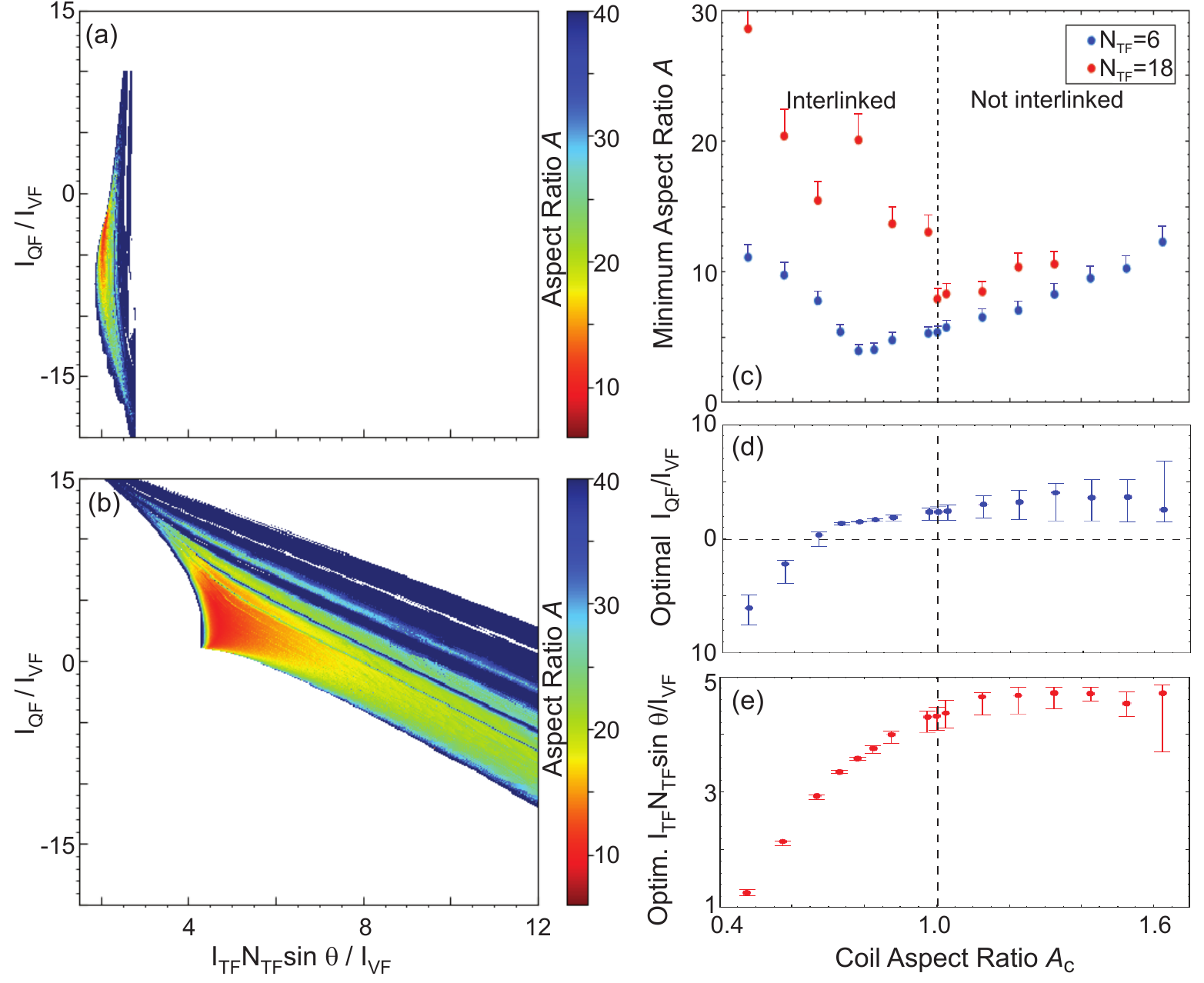} 
    \caption{(a)-(b) Like Fig.~\ref{fig:Cont} ($N$=6, $\theta=45^\circ$,
      $A_c$=0.67), but for different values of the coil aspect ratio
      ($A_c=0.575$ and $A_c=1.425$, respectively),
      all the rest remaining the same. 
      (c-e) Like Fig.~\ref{fig:NDepend}c-e, but as function of $A_c$.} 
    \label{fig:AcDepend}
  \end{center}
\end{figure}

\begin{figure}[t]
  \begin{center}
    \includegraphics{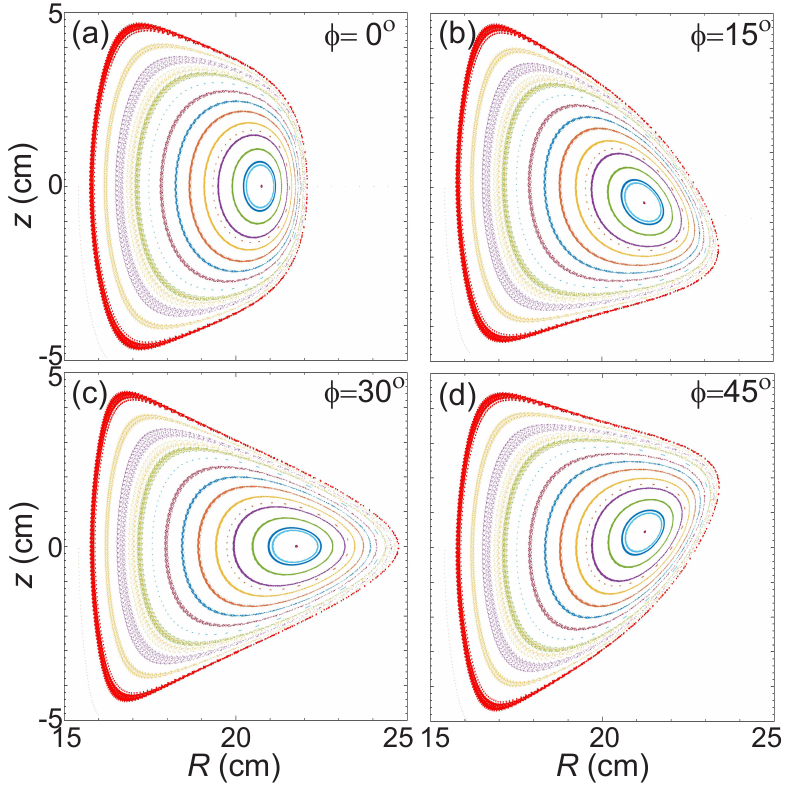} 
    \caption{Like Fig.~\ref{fig:Cont}, but for the best $N$=6 case in
    Fig.~\ref{fig:AcDepend}c. The plasma aspect ratio is $A$=4.0.}
    \label{fig:ContBestN6}
  \end{center}
\end{figure}

\begin{figure}[t]
  \begin{center}
    \includegraphics{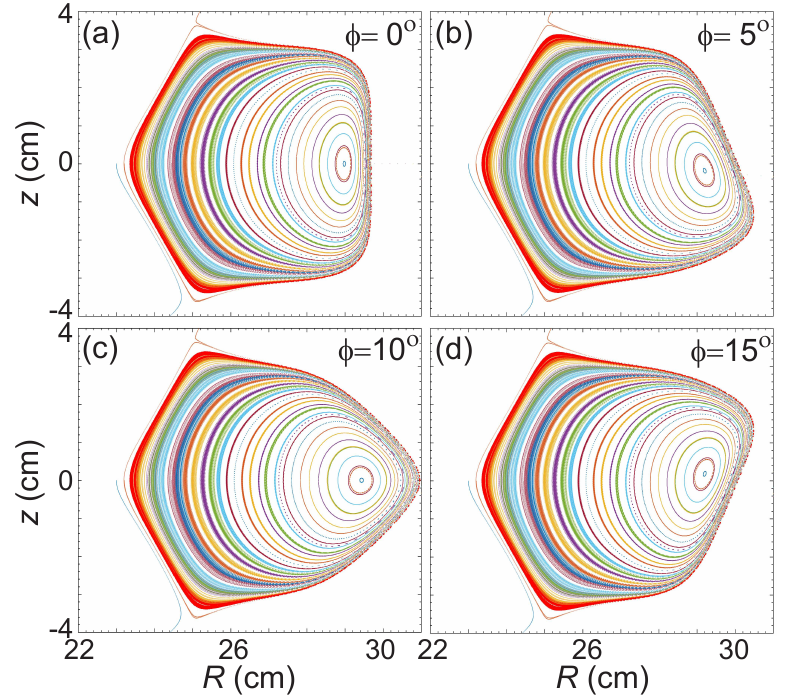} 
    \caption{Like Fig.~\ref{fig:Cont}, but for the best $N$=18 case in
    Fig.~\ref{fig:AcDepend}c. The plasma aspect ratio is $A$=7.9.}
    \label{fig:ContBestN18}
  \end{center}
\end{figure}

Fig.~\ref{fig:AcDepend} summarizes the results of the $A_c$ scan for
$N=6$. Of particular interest is that the largest plasma volumes (lowest
aspect ratio) are obtained for $A_c\lesssim 1$ (moderately interlinked coils),
for $N=6$, whereas for $N=18$ the lowest $A$ is obtained for $A_c=1$
(marginally interlinked coils). The corresponding flux surfaces are plotted
respectively in Fig.~\ref{fig:ContBestN6} and \ref{fig:ContBestN18}. 

\begin{figure}[t]
  \begin{center}
    \includegraphics{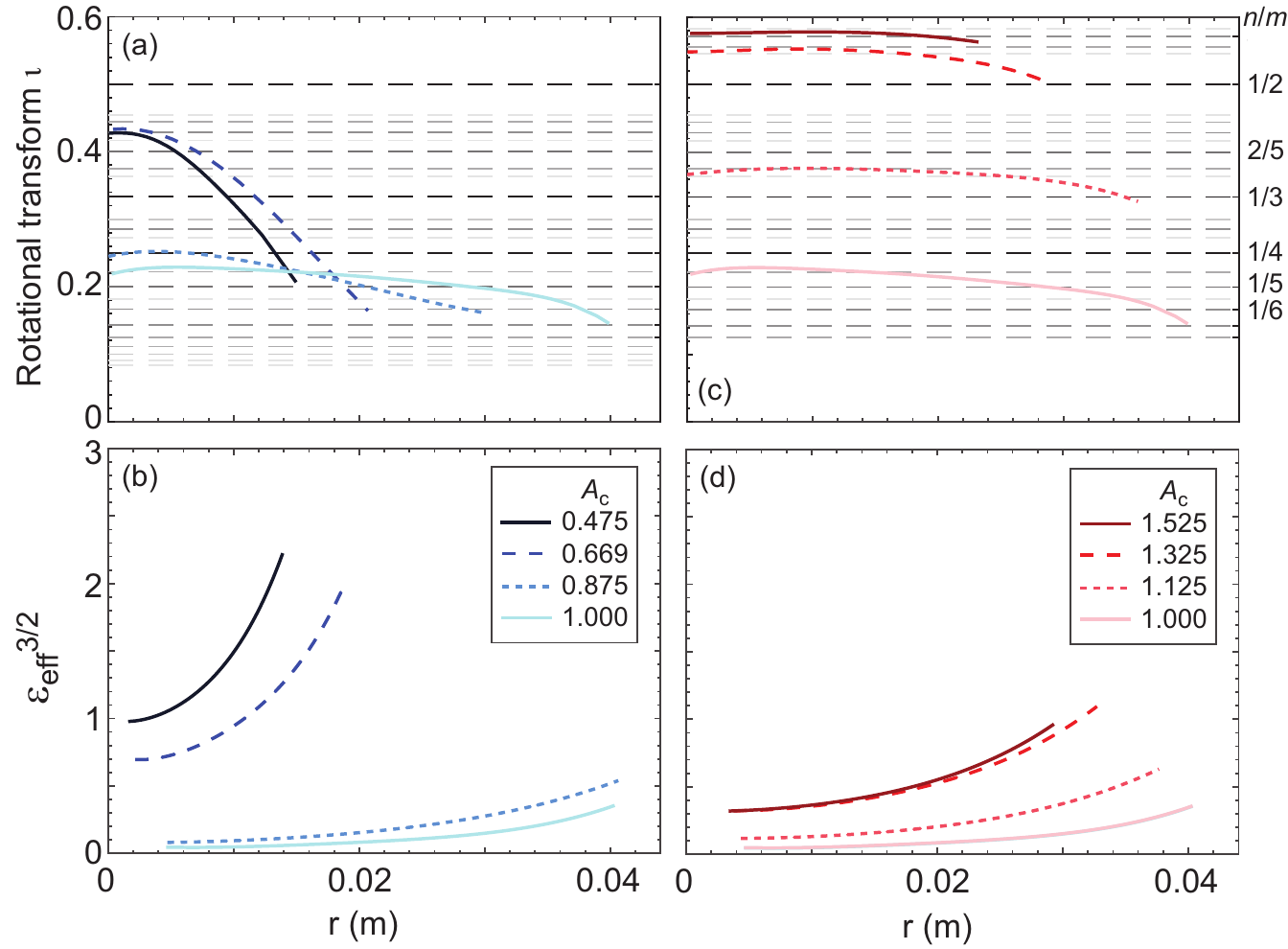} 
    \caption{Radial profiles of rotational transform and effective
      ripple for $N$=6 and for various coil aspect-ratios examined 
      in Fig.~\ref{fig:AcDepend}, grouped by (a-b) $A_c \le 1$
      (interlinked coils) and (c-d) $A_c \ge 1$ (non-interlinked coils). 
      }
    \label{fig:AcDependProfiles}
  \end{center}
\end{figure}

In addition, $A_c=1$ yields the lowest effective ripple
(Fig.~\ref{fig:AcDependProfiles}b and d)
but also one of the lowest rotational transform 
(Fig.~\ref{fig:AcDependProfiles}a and c). 

Note that, for any given $A_c$, there is a theoretical minimum below which
$A$ cannot be reduced. This is because the
aspect ratio of the plasma is 
necessarily larger than the aspect ratio of the coil-winding surface
($A>A_{CWS}$), which is related to $A_c$ as follows.
For $A_c<1$ the tilted coils are Villarceau circles for a toroidal
surface spanning $a_c-R_c \le R \le a_c+R_c$, hence the coil-winding surface
has major radius $R_{CWS}=a_c$, minor radius $a_{CWS}=R_c$ and aspect ratio
$A_{CWS}=1/A_c$. When instead $A_c>1$, it is simply 
$R_{CWS}=R_c$, $a_{CWS}=a_c$ and $A_{CWS}=A_c$. These lower limits
are not plotted in Fig.~\ref{fig:AcDepend}c, partly for simplicity and partly
because they are quite small: for $A_c=0.4$-1.7 they vary in the range 1-2.5.

\section{Discussion and conclusions}
%
As noted above, tilted-coil configurations are effectively
torsatrons of $m=n=1$. However, the present study might have
implications for helical devices in general:  
it is speculated that planar coils can realize stellarator
equilibria as well. It is also speculated that optimized configurations exist,
whose coils are more tilted and ``more planar'' 
than typical modular coils in optimized stellarators (in general less
coil-shaping implies less rotational transform, but this is compensated
for by increased coil-tilt, as per Fig.~\ref{fig:TiltDependProfiles}). 
In fact, some planar tilted coils are already used in W7-X, 
as mentioned in Sec.~\ref{subsec:rev}. 

A possible metric of
non-planarity is the root-mean-square deviation of the coil from a
plane, normalized to the coil diameter or perimeter. Its minimization
could be incorporated in the set of stellarator optimization criteria,
with a relative weight that will depend on coil-manufacturing times
and costs. 

This might seem in contrast with the complexity of the coil-winding surface 
(CWS) in W7-X, HSX and other optimized stellarators.
However, arbitrary current-patterns on an arbitrary CWS 
can always be approximated with planar current-filaments belonging to 
multiple {\em planes} intersecting the CWS. A high
enough number of adequately inclined planes should approximate
any current-pattern. If this results in coil-intersections, the
intersecting coils can be slightly displaced with respect to each other
in the minor radius direction (equivalent to introducing a second, concentric
CWS). Alternatively, the interesections can be removed in the same way as
intersecting TF and helical coils are replaced by modular coils.
In other words, the coils can be piecewise planar. 

To summarize and conclude, a numerical field-line tracer was used here to
compute the vacuum flux-surfaces generated by a variable number $N$ of
toroidal field coils, tilted by a variable angle $\theta$. 
Various normalized coil locations $A_c$ were also considered,
defined as the ratios between the 
major radius at which the coils are located, and the coil radii.
It was found that, for a particular geometry 
($N=6$, $\theta=45^\circ$and $A_c=0.78$, which can probably be optimized even
further) and coil-currents
($I_{QF}/I_{VF}=1.51$ and $I_{TF}/I_{VF}=0.84$),
tilted coil configurations can confine relatively large plasmas,
of aspect ratio as low as $A=4$.

Only vacuum flux-surfaces were computed in the present study,
to enable high-resolution scans of these and other parameters (for instance,
the coil-currents).
The results can be trusted in the low beta, $I_p$=0 limit.
Based on other works available in the literature 
\cite{Moroz_FST96,Moroz_PoP96,Moroz_NF97,Moroz_PPCF98,Moroz_PLA98}, 
including ours \cite{Clark14}, 
it is expected that finite bootstrap current and/or of a 
finite induced or driven $I_p$ should lead to an even lower $A$, which is left
as future work.

\section*{Acknowledgments}
The authors thank W.~Reiersen and D.~Spong (ORNL) for the encouragement
and fruitful discussions, as well as E.~Maragkoudakis (TU Eindhoven) and 
M.~Werl (TU Wien) for carefully reading and discussing the manuscript.

\section*{References}
\bibliographystyle{unsrt} 

\end{document}